\documentclass[12pt,a4paper]{article}

\usepackage[utf8]{inputenc}
\usepackage[english]{babel}
\usepackage[T1]{fontenc}
\usepackage{amsmath,amssymb,amsfonts,mathtools}
\usepackage{graphicx}
\usepackage{booktabs}
\usepackage{array,multirow}
\usepackage{float}
\usepackage{caption}
\usepackage{subcaption}
\usepackage{hyperref}
\usepackage{url}
\usepackage{xcolor}
\usepackage{geometry}
\usepackage{enumitem}
\usepackage{fancyhdr}
\usepackage[numbers,sort&compress]{natbib}
\newcommand{\MG}{\textsc{Mad\-Graph5\_aMC@NLO}}
\newcommand{\PY}{\textsc{Py\-thia}}

\title{Event-Level Classification of Boosted Multi-Tau Signatures in scNMSSM Di-Higgs Production}
\author{Marwa Telba}
\date{}

\begin{document}

\maketitle


\begin{abstract}
In the semi-constrained NMSSM at large $\lambda$ and low $\tan\beta$, di-Higgs production can proceed through $h_1h_1 \to 4a_1 \to 8\tau$ when the lightest pseudoscalar lies below the $b\bar{b}$ threshold. The $\tau$ pairs from each $a_1$ decay are collimated and merge within a single jet. Using the benchmark point $m_{a_1} = 8.2$~GeV identified in earlier work as a representative signal topology, we examine which observables separate such events from Standard Model processes producing similar final states. The exotic branching fraction of this point exceeds present experimental limits, and all results concern the shape of the distributions, not the signal rate. The substructure observables normally used for two-prong tagging prove ineffective here: the jet mass and the $N$-subjettiness ratio $\tau_{21}$ have separation powers of 0.02 and 0.09 against boosted $Z\to\tau^+\tau^-$. Both are ditau jets, and the neutrinos from $\tau$ decays displace the reconstructed masses to a common value near 18~GeV despite parent masses differing by an order of magnitude. Discrimination comes instead from the event as a whole. A boosted decision tree trained on single fat jets reaches an AUC of $0.759 \pm 0.002$, while an event-level classifier trained on 
complete events reaches $0.941 \pm 0.001$, driven by the multiplicity of boosted objects and the masses of the subleading ones. Truth matching confirms that these jets correspond to genuine $a_1$ decays in 77\% of events. Top-quark pair production 
reproduces the multi-object structure, and against $t\bar{t}\to b\bar{b}\tau^+\tau^-$ the 
AUC falls to $0.855 \pm 0.008$. For a cross-section-weighted combination of the two 
simulated backgrounds the classifier reaches $0.908 \pm 0.002$, with 71\% signal efficiency at 10\% background efficiency. These are classification metrics obtained from fast simulation, not discovery significances: reducible backgrounds, trigger acceptance and pile-up are not included.
\end{abstract}

\section{Introduction}
\label{sec:introduction}

One of the most challenging experimental problems in Higgs boson phenomenology is the identification of boosted di-tau systems at the Large Hadron Collider (LHC). Standard tau-tagging algorithms are optimised for resolved, isolated hadronic tau decays~\cite{CMS:2022mhs}, and lose efficiency when two tau leptons are produced in close proximity, a topology that arises naturally in models with light pseudoscalar states decaying as $a_1 \to \tau^+\tau^-$. The problem is not confined to new physics: in measuring Standard 
Model $H\to\tau^+\tau^-$ at high transverse momentum, the CMS Collaboration found it necessary to develop a dedicated reconstruction algorithm for the overlapping decay products~\cite{CMS:2024boosted}.

Higgs pair production is the primary route to the Higgs self-coupling at the LHC~\cite{LHCHiggsCrossSectionWorkingGroup:2016ypw}, and in extended scalar sectors it acquires decay channels absent in the Standard Model. In the Next-to-Minimal Supersymmetric Standard Model (NMSSM), the lightest CP-odd pseudoscalar $a_1$ can be sufficiently light to be produced copiously from the decays of heavier states, including the SM-like Higgs boson $h_1$~\cite{Ellwanger:2009dp, Curtin:2013fra}. 
When $m_{a_1} < 2m_b$, the decay $a_1 \to \tau^+\tau^-$ dominates, and if $a_1$ is produced with a large Lorentz boost the resulting $\tau$ pair is highly collimated with angular separation $\Delta R(\tau,\tau) \ll 1$. This boosted ditau topology was first studied by Katz, Son and Tweedie~\cite{Katz:2010iq}, who introduced the concept of a \textit{ditau-jet}, a single large-radius jet containing a merged 
$\tau^+\tau^-$ pair, and demonstrated its potential as a discovery channel at the LHC. Conte et al.~\cite{Conte:2016zpd} subsequently applied boosted ditau tagging to light NMSSM pseudoscalar states from superpartner cascades, demonstrating sensitivity at $\sqrt{s} = 13$~TeV with 20--50~fb$^{-1}$. Cacciapaglia et al.~\cite{Cacciapaglia:2017ofl} extended this to composite Higgs models, proposing a dedicated search for boosted ditau resonances for pseudoscalar masses below 65~GeV.

Multi-tau final states from di-Higgs production have also been searched for directly. The CMS Collaboration has set limits on $HH \to \tau\tau\tau\tau$ using 138~fb$^{-1}$~\cite{CMS:2023hh4tau}, but that search targets two Higgs bosons each decaying to a resolved $\tau$ pair. The topology considered here is different in both multiplicity and structure: $h_1h_1 \to 4a_1 \to 8\tau$ produces eight $\tau$ leptons arranged in four collimated pairs, and no existing analysis is optimised for it.

In our previous work~\cite{mtelba2026,mtelba2026b} we identified a benchmark point (BP2, $m_{a_1} = 8.2$~GeV) in the scNMSSM at large-$\lambda$/low-$\tan\beta$ in which this decay chain produces a distinctive boosted signature. A detector-level simulation showed that 82.4\% of same-$a_1$ $\tau$-pairs have $\Delta R(\tau,\tau) < 0.4$, below the standard jet cone radius, causing the pairs to merge within a single reconstructed jet. This follows from the large Lorentz boost $\gamma \approx m_{h_1}/(2m_{a_1}) \approx 7.5$. That work suggested that dedicated substructure taggers, exploiting the $a_1$ mass and 
$N$-subjettiness, might recover the sensitivity lost to this merging.

In this work we test that suggestion directly. Using the large-radius jets reconstructed by the detector simulation, we examine the full set of available substructure observables. The jet mass and the $N$-subjettiness ratio $\tau_{21}$ are the quantities most commonly used to identify two-prong objects~\cite{Larkoski:2017jix}. Neither provides useful separation from boosted $Z\to\tau^+\tau^-$. Both processes produce a collimated $\tau$ pair from a two-body decay, and the neutrinos carried away in $\tau$ decays displace the reconstructed mass far enough from the parent mass that an 8.2~GeV resonance and a 
91~GeV one become difficult to distinguish.

We then show that the discriminating information lies instead at the level of the event. Since $h_1h_1 \to 4a_1$ produces four boosted objects rather than one, the multiplicity of low-mass boosted objects and 
the masses of the subleading ones separate the signal from $Z\to\tau^+\tau^-$ considerably better than any single-jet observable. We further examine $t\bar{t}$ production, which reproduces much of this multi-object structure, and identify the $\tau$ multiplicity and missing transverse energy as the quantities that distinguish the signal in that case.

The paper is organised as follows. Section~\ref{sec:signal} describes the signal process and the benchmark point. Section~\ref{sec:background} presents the two simulated background processes. Section~\ref{sec:fatjets} details the fat jet reconstruction, and Section~\ref{sec:observables} examines the substructure observables individually. Section~\ref{sec:results} presents the classifier performance at both jet and event level. Section~\ref{sec:discussion} interprets these results and sets out the limitations of the study, and Section~\ref{sec:conclusion} summarises our findings.

\section{Signal Process and Benchmark Point}
\label{sec:signal}

\subsection{The benchmark point}

The signal process studied in this work is di-Higgs production in the scNMSSM at large-$\lambda$ and low-$\tan\beta$, focusing on the benchmark point BP2 identified in Ref.~\cite{mtelba2026b}. BP2 is characterised by a light CP-odd pseudoscalar with mass $m_{a_1} = 8.2$~GeV, below the $b\bar{b}$ 
threshold, so that the dominant decay is $a_1 \to \tau^+\tau^-$ with $\text{BR}(a_1 \to \tau^+\tau^-) = 85.1\%$. The di-Higgs production cross section at $\sqrt{s} = 13.6$~TeV is $\sigma(gg \to h_1 h_1) = 30.26$~fb, with Higgs self-coupling modifier $\kappa_\lambda = 0.907$. Since $\text{BR}(h_1 \to a_1 a_1) = 99.6\%$, the decay chain is essentially

\begin{equation}
gg \to h_1 h_1 \to 4a_1 \to 8\tau .
\end{equation}

This benchmark is used as a representative signal topology and not as a viable point of the model. With $\text{BR}(h_1 \to a_1 a_1) = 99.6\%$ the total width of $h_1$ is 1.2~GeV, about 300 times the Standard Model value, and the decays that remain visible are suppressed by the same factor. Enabling the LHC Higgs constraints of \textsc{NMSSMTools} (\texttt{MODSEL 19 = 1}) flags the point as excluded by searches for $H\to AA$ in $\mu$, $\tau$ and $b$ final states, and reports an undetected branching fraction of 99.6\%, against present limits of 12--18\% on undetected non-standard Higgs decays~\cite{ATLAS:HiggsMap2022,ATLAS:HiggsComb2026}.. The quantities studied below are shape-based and do not depend on the overall signal rate, as discussed in Section~\ref{sec:discussion}.

The defining feature of this benchmark is the large Lorentz boost of the $a_1$,

\begin{equation}
\gamma \approx \frac{m_{h_1}}{2m_{a_1}} 
\approx 7.5,
\end{equation}

which collimates the $\tau$ pair from each $a_1$ decay to a characteristic angular separation

\begin{equation}
\Delta R(\tau,\tau) \approx 
\frac{2}{\gamma} \approx 0.27,
\end{equation}

well inside the standard jet cone radius of $R = 0.4$. The benchmark parameters are collected in Table~\ref{tab:bp2}.

\begin{table}[h!]
\centering
\renewcommand{\arraystretch}{1.3}
\caption{Parameters of benchmark point BP2. 
Masses, couplings and branching ratios are taken 
from Ref.~\cite{mtelba2026b}; the boost and 
angular separation follow from them.}
\label{tab:bp2}
\begin{tabular}{lc}
\hline\hline
Parameter & Value \\
\hline
$m_{h_1}$ [GeV]                        & 123.3 \\
$m_{a_1}$ [GeV]                        & 8.2   \\
$\kappa_\lambda$                       & 0.907 \\
$\sigma(gg\to h_1h_1)$ [fb]            & 30.26 \\
$\text{BR}(h_1\to a_1a_1)$ [\%]        & 99.6  \\
$\text{BR}(a_1\to\tau^+\tau^-)$ [\%]   & 85.1  \\
$\text{BR}(a_1\to gg)$ [\%]            & 14.4  \\
Lorentz boost $\gamma$                 & 7.5   \\
$\Delta R(\tau,\tau)$ [typical]        & 0.27  \\
Same-$a_1$ pairs with $\Delta R < 0.4$ & 82.4\% \\
\hline\hline
\end{tabular}
\end{table}

The inclusive cross section for the complete $8\tau$ final state is

\begin{align}
\sigma(8\tau) &= \sigma(gg\to h_1h_1) 
\times [\text{BR}(h_1\to a_1a_1)]^2 
\times [\text{BR}(a_1\to\tau^+\tau^-)]^4 
\nonumber\\
&\approx 30.26 \times (0.996)^2 \times 
(0.851)^4 \approx 15.7~\text{fb},
\end{align}

where all four $a_1$ bosons are assumed to decay to $\tau^+\tau^-$. This is the dominant contribution, the remainder involving at least one $a_1 \to gg$ decay [$\text{BR}(a_1\to gg) = 14.4\%$]. The figure characterises the size of the signal but is not used in what follows: the classifier performance reported in Section~\ref{sec:results} is expressed in terms of relative efficiencies, and no absolute event yields are quoted.

\subsection{Signal simulation}

The signal sample was generated with \textsc{MadGraph5\_aMC@NLO}~\cite{Alwall:2014hca} at leading order with the \texttt{NNPDF2.3LO} parton distribution functions~\cite{Ball:2012cx}, at $\sqrt{s} = 13.6$~TeV. Parton showering and hadronisation were performed with \textsc{Pythia}~8.316~\cite{Bierlich:2022pfr}, 
and the detector response with \textsc{Delphes}~3~\cite{deFavereau:2013fsa}. A total of $10^5$ events were produced.

The $a_1$ is not part of the Standard Model particle content, and the branching ratios computed by \texttt{NMSSMTools} are not propagated automatically through the simulation chain. The decay tables for $h_1$ and $a_1$ were therefore written explicitly into the Les Houches event file before showering, following the procedure of Ref.~\cite{mtelba2026b}. Validation on a subsample confirmed a mean of 7.6 $\tau$ leptons per event, consistent with the eight expected from the decay chain after accounting for the subleading $a_1 \to gg$ mode.

The same event sample underlies Ref.~\cite{mtelba2026b}, where the detector simulation showed that 82.4\% of $\tau$ pairs originating from a common $a_1$ satisfy $\Delta R(\tau,\tau) < 0.4$ and therefore merge 
within a single reconstructed jet. The detector step was repeated for the present study with a modified fat jet configuration, described in Section~\ref{sec:fatjets}.

\section{Background Samples}
\label{sec:background}
Two background processes were simulated. Both were generated at leading order with 
\MG~\cite{Alwall:2014hca}, using the \texttt{NNPDF2.3LO} parton distribution 
functions~\cite{Ball:2012cx}. Showering was performed with \PY~8.316~\cite{Bierlich:2022pfr}, and the events were passed through the same 
detector configuration as the signal.

\subsection{\texorpdfstring{$Z/\gamma^*\to\tau^+\tau^-$}{Z/gamma* to tau+ tau-}}
Inclusive $Z$ production is copious, with a cross section of order 60~nb at $\sqrt{s}=13.6$~TeV and 
$\text{BR}(Z\to\tau^+\tau^-)=3.37\%$~\cite{ParticleDataGroup:2024cfk}, but most events yield well-separated $\tau$ pairs with $\Delta R(\tau,\tau)\sim 2$--$3$ and are trivially distinguished from the signal. The 
relevant component is the boosted one, in which the $\tau$ pair is collimated in the same way as in $a_1\to\tau^+\tau^-$. We therefore generated $p p \to Z(\to\tau^+\tau^-)$ and $pp \to Z(\to\tau^+\tau^-)j$ with a generation-level requirement $p_T^{\tau\tau} > 100$~GeV, obtaining $\sigma = 18.35 \pm 0.02$~pb for 
$5\times 10^4$ events. No matrix-element to parton-shower merging was applied; at leading order the $2\to1$ process has $p_T^{\tau\tau} = 0$ and is removed entirely by the generation-level requirement, so the sample 
is effectively $Z+\text{jet}$ production.

\subsection{$t\bar{t}$}

Top-quark pair production is the second process considered. The inclusive cross section is large, $\sigma_{t\bar{t}} = 850 \pm 27$~pb 
at $\sqrt{s} = 13.6$~TeV~\cite{ATLAS:2024ttbar}, but only the fraction in which both $W$ bosons decay to $\tau\nu$ produces a final state 
resembling the signal. We generated 
\begin{equation}
pp \to t\bar{t},\quad 
t \to W^+ b \to \tau^+\nu_\tau b,\quad
\bar{t} \to W^- \bar{b} \to \tau^-\bar{\nu}_\tau \bar{b},
\end{equation}
obtaining $\sigma = 5.68 \pm 0.01$~pb for $10^4$ events. The decay chain was fixed at matrix-element level, so this figure already includes both $W \to \tau\nu$ branching fractions; the subsequent $\tau$ decays are handled by \textsc{Pythia} and are not folded into the quoted cross section. No $K$-factor was applied, and the value is the generator cross section at leading order. For comparison, scaling the measured inclusive cross section by $[\text{BR}(W\to\tau\nu)]^2 = (0.1138)^2$ gives 11.0~pb at NNLO, which our leading-order value reproduces with an effective $K$-factor of 1.9, consistent with expectations for this process.

This process deserves attention because it produces four hard objects per event (two $b$-jets and two $\tau$ leptons) together with substantial missing transverse energy from the four neutrinos of the $W$ and subsequent $\tau$ decays. It therefore reproduces, at least superficially, the multi-object structure that 
characterises $h_1h_1\to 4a_1$.

\subsection{Processes not included}

Several contributions are absent from this study. Multi-jet production with jets misidentified as $\tau$ candidates cannot be estimated reliably from fast simulation, since fake rates depend strongly on detector conditions and are normally obtained from data. $W$+jets, single-top and diboson production were likewise not simulated. The results below should therefore be understood as a comparison against the two simulated processes, not as a complete background estimate.

\section{Fat Jet Reconstruction}
\label{sec:fatjets}

\subsection{Jet Definition}

The analysis of Ref.~\cite{mtelba2026b} employed anti-$k_t$ jets~\cite{Cacciari:2011ma} with $R = 0.4$ and found that the majority of same-$a_1$ $\tau$-pairs fall within a single such cone. To reconstruct the merged system as a whole, we use the large-radius jets provided by the \textsc{Delphes}~3 \texttt{FatJetFinder} module, which clusters particle-flow objects with the anti-$k_t$ algorithm~\cite{Cacciari:2008gp} at $R = 0.8$ and computes $N$-subjettiness, trimming, pruning and soft-drop observables internally.

The radius $R = 0.8$ is large enough to contain both $\tau$ decay products for the majority of boosted $a_1$ decays, while remaining small enough to limit contamination from the other $a_1$ bosons in the same event. Truth-level matching, described in 
Section~\ref{sec:results}, confirms this: only 11.2\% of reconstructed fat jets lie within $\Delta R < 0.8$ of more than one $a_1$.

This radius is a baseline reconstruction choice rather than an experimentally optimised value, as is the 30~GeV threshold introduced below. Both would need revisiting in a full analysis: a low-$p_T$ large-radius jet is particularly sensitive to pile-up, and the trigger considerations discussed in Section~\ref{sec:discussion} apply directly to this configuration.

\subsection{Transverse Momentum Threshold}

The default \textsc{Delphes} CMS card applies $p_T^{\rm min} = 200$~GeV to fat jets, a value tuned for boosted electroweak bosons and top quarks. This threshold is inappropriate for the present signal: with $m_{a_1} = 8.2$~GeV the boosted objects are comparatively soft, and at 200~GeV only 18\% of BP2 events contain any fat jet at all, with a mean multiplicity of 0.18 per event.

We therefore lower the threshold to $p_T^{\rm min} = 30$~GeV, keeping all other card 
settings unchanged, and re-run the detector simulation for both signal and background. This threshold is a reconstruction-level configuration rather than a property of the 
detector, and lowering it is equivalent to extending the kinematic range of an analysis; the calorimeter and tracking parametrisations are left untouched. The effect is substantial and is summarised in Table~\ref{tab:ptmin}.

\begin{table}[h!]
\centering
\renewcommand{\arraystretch}{1.3}
\caption{Effect of the fat jet $p_T$ threshold on 
BP2 signal acceptance.}
\label{tab:ptmin}
\begin{tabular}{lcc}
\hline\hline
Quantity & $p_T^{\rm min}=200$~GeV 
         & $p_T^{\rm min}=30$~GeV \\
\hline
Fat jets per event           & 0.18  & 3.55  \\
Events with $\geq 1$ fat jet & 18\%  & 99.5\% \\
Events with $\geq 2$ fat jets & --   & 95.5\% \\
Median $m_J$ [GeV]           & 60.9  & 13.4  \\
Median $p_T(J)$ [GeV]        & 286.6 & 60.9  \\
\hline\hline
\end{tabular}
\end{table}

At the lower threshold the mean multiplicity reaches 3.55 fat jets per event. The decay chain contains four parent $a_1$ bosons, which give rise to several reconstructed low-mass fat jets; the two numbers need not coincide, since some $a_1$ fall outside acceptance and others merge.

\subsection{Selection}

Fat jets are reconstructed with $p_T(J) > 30$~GeV, as set in the detector card. No additional pseudorapidity requirement is imposed at analysis level; the acceptance is that of the particle-flow objects from which the jets are built. No truth-level matching is applied: all reconstructed fat jets are retained, so that the classifier operates on the same information that would be available in data.

Alongside the fat jets we count $\tau$-tagged small-radius jets. We define $N_\tau$ as the number of $R = 0.4$ jets with $p_T > 20$~GeV and $|\eta| < 2.5$ carrying the \textsc{Delphes} tau-identification flag, which corresponds to hadronically decaying $\tau$ candidates. The quantity is reconstructed-level and uses no truth information. No overlap removal is applied between these jets and the fat jets, so a $\tau$-tagged jet lying inside a fat jet contributes to both.

\section{Fat Jet Observables}
\label{sec:observables}

\subsection{Observables Available in the Simulation}

The \texttt{FatJetFinder} module computes a set of substructure quantities for every fat jet. We consider all of them, together with the basic kinematic variables:

\begin{itemize}
\item Jet mass $m_J$ and transverse momentum $p_T(J)$.
\item $N$-subjettiness $\tau_1$, $\tau_2$, $\tau_3$~\cite{Thaler:2010tr,Thaler:2012sz}, computed with the angular weighting exponent $\beta = 1$, the \textsc{Delphes} default and the value conventionally used for two-prong tagging, and the ratios $\tau_{21} = \tau_2/\tau_1$ and $\tau_{32} = \tau_3/\tau_2$.

\item The soft-drop mass $m_{\rm SD}$ and the number of subjets identified by the soft-drop~\cite{Larkoski:2014wba}, trimming~\cite{Krohn:2009th} and pruning~\cite{Ellis:2009me} algorithms 
      ($N_{\rm SD}$, $N_{\rm trim}$, $N_{\rm prun}$).
\item The energy sharing between the two soft-drop subjets,
      \begin{equation}
      z_{\rm sub} = 
      \frac{\min(E_1, E_2)}{E_1 + E_2},
      \end{equation}
      where $E_1$ and $E_2$ are the energies of the two subjets identified by the soft-drop algorithm. This quantity approaches 0.5 for a symmetric two-prong decay and takes small values for a single hard core accompanied by soft radiation.
\item The subjet asymmetry,
      \begin{equation}
      A_{\rm sub} = 
      \frac{|E_1 - E_2|}{E_1 + E_2}
      = 1 - 2z_{\rm sub},
      \end{equation}
      which is retained for completeness although it carries no information beyond $z_{\rm sub}$.      
\item Charged and neutral constituent multiplicities and energy fractions.
\end{itemize}

Energy correlation functions~\cite{Larkoski:2013eya} are not among the quantities stored by the simulation and are not used here.

\subsection{Separation Power}
To assess each observable individually before combining them, we use the standard measure
\begin{equation}
S = \frac{|\langle x\rangle_{\rm sig} - 
           \langle x\rangle_{\rm bkg}|}
         {\sqrt{\tfrac{1}{2}
          (\sigma_{\rm sig}^2 + 
           \sigma_{\rm bkg}^2)}},
\end{equation}

evaluated over all reconstructed fat jets in the two samples. Values of $S$ for the eleven observables examined are listed in Table~\ref{tab:sep}.

\begin{table}[h!]
\centering
\renewcommand{\arraystretch}{1.3}
\caption{Separation power $S$ of individual 
fat jet observables between the BP2 signal and 
the boosted $Z\to\tau^+\tau^-$ background. 
Means are taken over all reconstructed fat jets 
($3.5\times10^5$ signal, $1.4\times10^5$ 
background).}
\label{tab:sep}
\begin{tabular}{lccc}
\hline\hline
Observable & Signal & Background & $S$ \\
\hline
$z_{\rm sub}$       & 0.217 & 0.091 & 0.359 \\
$N_{\rm SD}$        & 1.93  & 1.75  & 0.340 \\
$N_{\rm prun}$      & 1.96  & 1.81  & 0.337 \\
$\tau_1$            & 0.189 & 0.149 & 0.331 \\
$N_{\rm trim}$      & 2.47  & 2.07  & 0.275 \\
$A_{\rm sub}$       & 0.431 & 0.321 & 0.240 \\
$p_T(J)$ [GeV]      & 83.1  & 97.1  & 0.205 \\
$m_{\rm SD}$ [GeV]  & 13.07 & 11.02 & 0.128 \\
$\tau_{21}$         & 0.542 & 0.560 & 0.090 \\
$\tau_{32}$         & 0.634 & 0.639 & 0.025 \\
$m_J$ [GeV]         & 18.28 & 18.54 & 0.016 \\
\hline\hline
\end{tabular}
\end{table}

These observables are not independent. The three subjet counts are computed by different grooming algorithms applied to the same jet, $m_J$ and $m_{\rm SD}$ differ only by the soft-drop procedure, the $N$-subjettiness ratios share a common numerator or denominator, and $A_{\rm sub}$ is an exact function of $z_{\rm sub}$. The feature importances quoted in Section~\ref{sec:results} should therefore not be read as assigning independent information to each input: when several correlated variables are available, a decision tree distributes the weight among them in a way that depends on the training sample.

\subsection{Interpretation}
Two features of Table~\ref{tab:sep} determine the strategy adopted in the remainder of this paper.
First, no single observable separates the two samples strongly. The largest value is $S = 0.36$ for $z_{\rm sub}$, corresponding to a shift of roughly one third of the combined width. A selection based on any one of these quantities will therefore be inefficient.
Second, $\tau_{21}$ and $m_J$, the two observables most commonly used to identify boosted two-prong objects and those suggested in Ref.~\cite{mtelba2026b}, are among the weakest, with $S = 0.09$ and $S = 0.02$ respectively.

This is a consequence of the background definition. The comparison is against $Z\to\tau^+\tau^-$ produced with $p_T^{\tau\tau} > 100$~GeV, so the background jet also contains a collimated $\tau$ pair from a two-body decay. Both samples consist of two-prong objects decaying to the same final state, and $\tau_{21}$ measures precisely the degree of two-prong structure. It therefore carries little information about which parent produced the pair.

The parent masses differ by an order of magnitude, 8.2~GeV against 91.2~GeV, and in the absence of neutrinos this would separate the two cleanly. The measured masses do not: $m_J = 18.3$~GeV for the signal and 18.5~GeV for the background. Each $\tau$ carries away at least one neutrino, so the reconstructed mass is the visible mass of the decay products rather than the parent mass. For the $Z$ this represents a large downward shift; for the $a_1$ the jet mass is instead increased by soft radiation clustered into the $R = 0.8$ cone. 
The two effects bring the distributions into near coincidence.

The observables that retain some separation are those sensitive to how energy is distributed within the jet rather than to its total invariant mass: $z_{\rm sub}$, the subjet counts, and $\tau_1$. We note that $\tau_1$ ($S = 0.33$) is considerably more informative than the ratio $\tau_{21}$ ($S = 0.09$), indicating that normalising $\tau_2$ by $\tau_1$ removes information relevant to this particular discrimination.

\begin{figure}[h!]
\centering
\includegraphics[width=\textwidth]
    {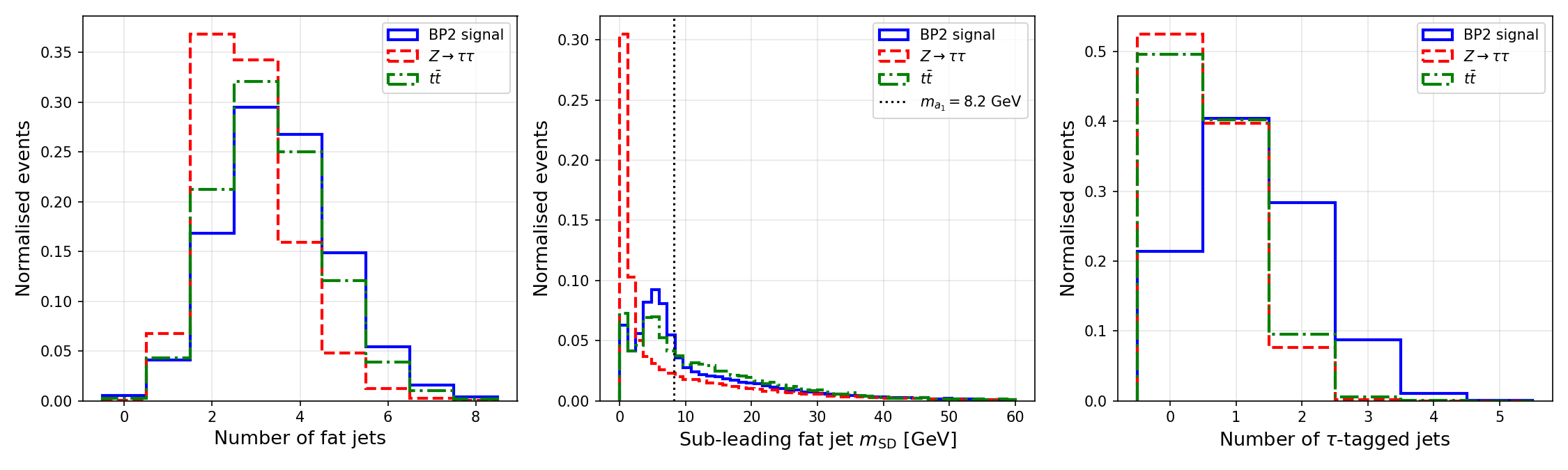}
\caption{Distributions for the BP2 signal (solid 
blue), boosted $Z\to\tau^+\tau^-$ (dashed red) 
and $t\bar{t}$ (dash-dotted green). Left: number 
of fat jets per event. Centre: soft-drop mass of 
the sub-leading fat jet, with the dotted line 
indicating $m_{a_1} = 8.2$~GeV. Right: number of 
$\tau$-tagged jets. All distributions are 
normalised to unit area.}
\label{fig:dist}
\end{figure}

The left panel of Fig.~\ref{fig:dist} points to the alternative followed in Section~\ref{sec:results}. The signal contains 3.55 fat jets per event on average, compared with 2.80 in $Z\to\tau^+\tau^-$: the decay chain $h_1h_1 \to 4a_1$ produces four 
parent bosons, whereas $Z\to\tau^+\tau^-$ produces one, the remaining jets arising from initial-state radiation. Whether this distinction survives against a background that also produces several hard objects is examined in Section~\ref{sec:results}.

\section{Results}
\label{sec:results}

\subsection{Single-jet classification}

Given the separation powers listed in Table~\ref{tab:sep}, no useful selection could be constructed from any single observable, and a multivariate combination was therefore adopted. We use the gradient-boosted decision tree of \textsc{scikit-learn}~\cite{scikit-learn} with 200 estimators, maximum depth three, learning rate 0.1, subsample fraction 0.8, the Friedman mean squared error splitting criterion, and a fixed random seed. These are the library defaults apart from the number of estimators and the subsample fraction; no hyperparameter search was performed, and the values reported below should be regarded as a baseline rather than an optimised limit. No preprocessing, scaling or feature selection was applied, since decision trees are invariant under monotonic transformations of the inputs.

Signal and background samples were subsampled to equal size to avoid a class imbalance, and the resulting set was divided 70/30 between training and testing. All figures of merit refer to the test partition.

At jet level every reconstructed fat jet forms an independent entry, so events contribute in proportion to their fat jet multiplicity. Since the signal averages 3.55 fat jets per event against 2.80 in the background, signal events carry roughly 27\% more weight than background events in this training. This is a further reason why the jet-level figures are not directly comparable with the event-level ones, which treat each event equally.

No truth matching is used in defining the training labels: every reconstructed fat jet in a signal event is labelled as signal, and every fat jet in a background event as background. At event level the label refers to the process rather than to the 
presence of a particular object. The matching rates quoted in Section~\ref{subsec:decomp} serve to interpret the result, not to construct it.

The classifier was first trained on individual fat jets against the $Z\to\tau^+\tau^-$ sample, using the eleven observables of Table~\ref{tab:sep}. This yielded an area under the ROC curve of $0.759 \pm 0.002$ in five-fold cross-validation, compared with 0.757 for the single representative split quoted in Table~\ref{tab:auc}, with $36.3 \pm 0.3$\% of signal jets retained at 10\% background efficiency. The two leading inputs were $m_{\rm SD}$ (0.378) and $p_T(J)$ (0.345), while $\tau_{21}$ contributed 0.085. Combining several weakly separating observables therefore improved on any one of them, but did not produce a classifier strong enough to support an analysis.

\subsection{Event-level classification}

The fat jet multiplicities noted at the end of Section~\ref{sec:observables} suggest a different approach. A separate gradient-boosted decision tree was therefore trained on complete events, with the same hyperparameters but a different input representation.

The inputs comprised the fat jet multiplicity, $H_T$, $E_T^{\rm miss}$ and the number of 
$\tau$-tagged $R = 0.4$ jets, followed by the nine observables for each of the three highest-$p_T$ fat jets and the angular separation between the two leading ones, giving 32 inputs in total. Events containing fewer than three fat jets had the missing entries set to zero. Against $Z\to\tau^+\tau^-$ this configuration gave an AUC of $0.941 \pm 0.001$, retaining $82.2 \pm 0.3$\% of signal events at 10\% background efficiency, $71.7 \pm 0.5$\% at 5\% and $47.9 \pm 1.8$\% at 1\%.

\subsection{Decomposition of the discrimination}
\label{subsec:decomp}

\begin{figure}[h!]
\centering
\includegraphics[width=0.7\textwidth]
    {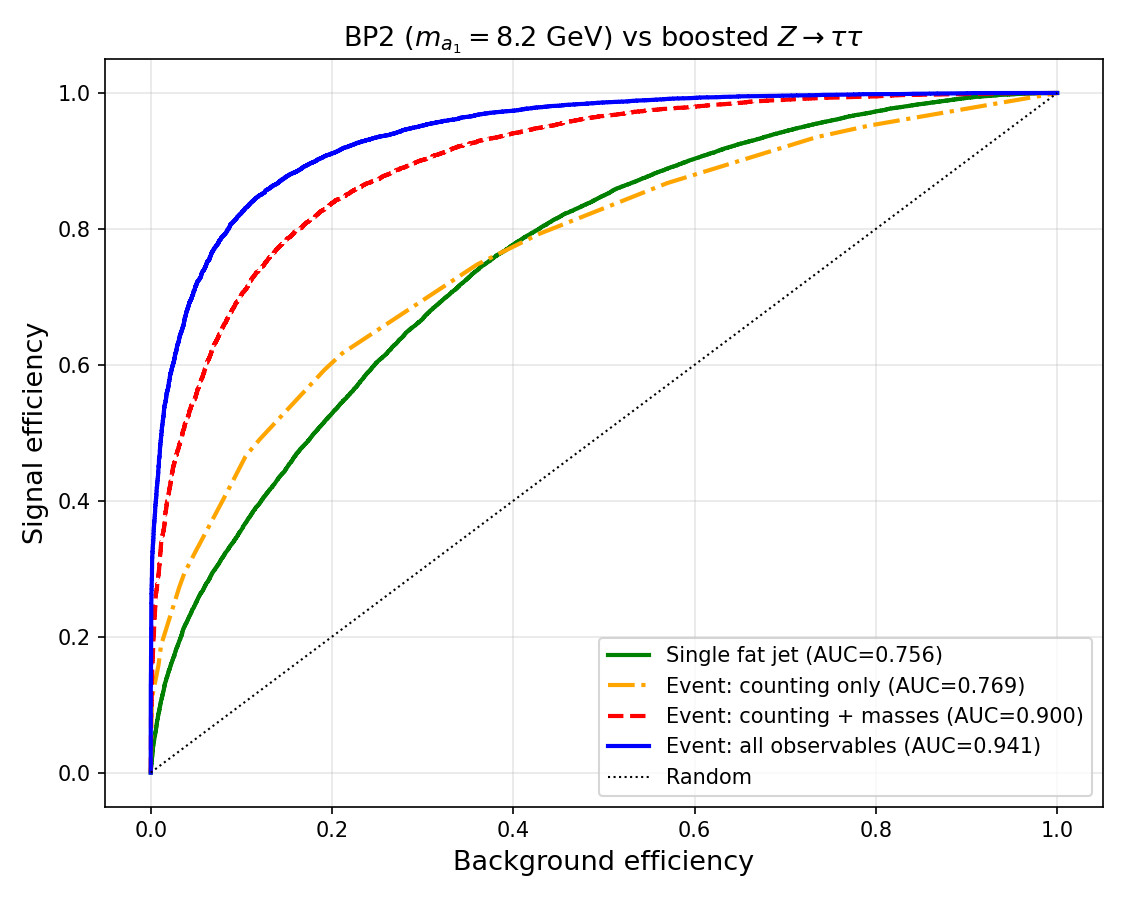}
\caption{ROC curves for the input sets listed in 
Table~\ref{tab:auc}. The dotted diagonal 
corresponds to random classification.}
\label{fig:roc}
\end{figure}

In order to identify the origin of this improvement, the training was repeated with 
restricted sets of inputs. The results are summarised in Table~\ref{tab:auc}, and the 
corresponding ROC curves are shown in Fig.~\ref{fig:roc}.

\begin{table}[h!]
\centering
\renewcommand{\arraystretch}{1.3}
\caption{Classifier performance against 
$Z\to\tau^+\tau^-$ for different input sets, from 
a single train/test split. $\varepsilon_S$ denotes 
the signal efficiency at the stated background 
efficiency $\varepsilon_B$. Cross-validated values 
for the first and fourth rows are given in 
Table~\ref{tab:cv}.}
\label{tab:auc}
\begin{tabular}{lccc}
\hline\hline
Input set & AUC 
  & $\varepsilon_S$ ($\varepsilon_B=10\%$) 
  & $\varepsilon_S$ ($\varepsilon_B=5\%$) \\
\hline
Single fat jet            & 0.757 & 0.359 & 0.248 \\
Event: $N_J$, $N_\tau$    & 0.769 & 0.464 & 0.330 \\
Event: multiplicity + $m_{\rm SD}$ 
                          & 0.900 & 0.703 & 0.555 \\
Event: all 32 inputs      & 0.941 & 0.823 & 0.717 \\
Event: no $p_T$, no $H_T$ & 0.936 & 0.807 & 0.701 \\
\hline\hline
\end{tabular}
\end{table}

What stands out in this table is that two counts alone, $N_J$ and $N_\tau$, already exceeded the single-jet classifier trained on eleven substructure observables. Supplementing these with the soft-drop masses of the three leading fat jets raised the AUC to 0.900, while the remaining inputs of the full configuration added a further 
0.041.

The feature importances of the full configuration support the same interpretation. The soft-drop masses of the second and third fat jets carried weights of 0.170 and 0.097 respectively, together more than eight times that of the leading jet (0.031). Discrimination against $Z\to\tau^+\tau^-$ was therefore driven not by the hardest object in the event, but by whether additional low-mass boosted objects accompanied it. These masses retain information about the presence and energy distribution of additional boosted activity, but they do not provide a direct reconstruction of $m_{a_1}$: the mean value for $j_2$, 13.6~GeV, lies well above 8.2~GeV. What the classifier uses is the pattern of several low-mass objects in the same event, not a mass peak. In the background the subleading jets originate from initial-state radiation and carry no relation to the mass of the leading jet.

This interpretation can be tested directly. Each of the three leading fat jets was matched to truth-level $a_1$ bosons with $p_T > 20$~GeV by requiring $\Delta R(J, a_1) < 0.8$. A jet counts as matched if at least one $a_1$ satisfies this condition; the assignment is not required to be one-to-one, and 11.2\% of jets lie within the matching radius of more than one $a_1$. Matching is used solely to interpret the classifier output and plays no part in defining the training labels.

The match rates are 68.4\%, 74.6\% and 55.5\% for $j_1$, $j_2$ and $j_3$ respectively, and 76.8\% of signal events contain at least two matched jets. That $j_2$ is matched more often than $j_1$ is consistent with the feature importances: the hardest jet is more frequently a merged or radiative object, while the subleading jets are cleaner single $a_1$ candidates. Matched jets are also systematically lighter than unmatched ones, 21.2~GeV against 37.3~GeV for $j_1$, confirming that the unmatched component is QCD activity rather than signal.

\subsection{Stability against the background selection}

The background sample was generated with $p_T^{\tau\tau} > 100$~GeV, a requirement that shapes not only its momentum spectrum but every quantity correlated with it. Two tests were performed to establish how far the reported performance depends on this choice.

Removing $H_T$ and all three fat jet $p_T$ inputs changed the AUC from 0.941 to 0.936, and the signal efficiency at 10\% background efficiency from 82.3\% to 80.7\%. The classifier is therefore not reading the momentum variables directly.

A second concern is the treatment of events with fewer than three fat jets, whose missing entries are set to zero. This could in principle allow the classifier to infer the multiplicity from the padding pattern itself. Adding three explicit binary masks, indicating whether the event contains at least one, two or three fat jets, to the 32 inputs left the weighted AUC unchanged at $0.9077 \pm 0.0020$. Removing $N_J$ while retaining the masks changed it only to $0.9069 \pm 0.0020$; the multiplicity remains available through the masks and through the populated feature blocks, so this test isolates the explicit count rather than the information itself.

This does not by itself establish independence from the generation cut, since the requirement also shapes the jet multiplicities and masses. We therefore generated a second $Z\to\tau^+\tau^-$ sample with $p_T^{\tau\tau} > 50$~GeV ($\sigma = 106.9 \pm 0.2$~pb, $2\times10^4$ events) and trained and tested across the two slices. The results are given in Table~\ref{tab:domain}.

\begin{table}[h!]
\centering
\renewcommand{\arraystretch}{1.3}
\caption{Cross-slice performance of the 
event-level classifier. Each entry gives the AUC 
and, in parentheses, the signal efficiency at 
10\% background efficiency.}
\label{tab:domain}
\begin{tabular}{lcc}
\hline\hline
& \multicolumn{2}{c}{Tested on} \\
Trained on & $p_T^{\tau\tau} > 100$~GeV 
           & $p_T^{\tau\tau} > 50$~GeV \\
\hline
$p_T^{\tau\tau} > 100$~GeV 
  & 0.936 (0.803) & 0.727 (0.372) \\
$p_T^{\tau\tau} > 50$~GeV  
  & 0.867 (0.620) & 0.910 (0.738) \\
\hline\hline
\end{tabular}
\end{table}

The response is strongly asymmetric. A classifier trained on the harder slice loses 0.209 in AUC when applied to the softer one, whereas training on the softer slice costs only 0.043 in the reverse direction. The softer sample is systematically less energetic in every relevant quantity: mean $H_T$ falls from 279 to 173~GeV, $p_T(j_1)$ from 145 to 87~GeV, and the fat jet multiplicity from 2.80 to 2.41. A model trained only on the harder slice has never encountered background of this kind and misclassifies part of 
it as signal.

Two conclusions follow. The discrimination is not an artefact of the generation cut, since a classifier trained on the broader slice retains an AUC of 0.867 when tested on the narrower one. But the values reported elsewhere in this paper apply within the kinematic range on which the classifier was trained, and should not be assumed to hold for softer background. An experimental analysis would need to train across the full 
$p_T$ range rather than on a slice of it.

\subsection{Top-quark background}

The interpretation above rests on the multiplicity of boosted objects, and $t\bar{t}$ production is expected to populate the same region. Table~\ref{tab:means} compares the two backgrounds with the signal.

\begin{table}[h!]
\centering
\renewcommand{\arraystretch}{1.3}
\caption{Mean values of selected event-level 
quantities.}
\label{tab:means}
\begin{tabular}{lccc}
\hline\hline
Quantity & Signal & $Z\to\tau^+\tau^-$ 
         & $t\bar{t}$ \\
\hline
$N_J$                     & 3.55  & 2.80  & 3.36  \\
$N_\tau$                  & 1.28  & 0.55  & 0.61  \\
$E_T^{\rm miss}$ [GeV]    & 56.3  & 67.7  & 78.7  \\
$H_T$ [GeV]               & 306.8 & 279.3 & 274.0 \\
$m_{\rm SD}(j_2)$ [GeV]   & 13.57 & 8.13  & 12.95 \\
$m_{\rm SD}(j_3)$ [GeV]   & 7.78  & 4.72  & 7.30  \\
\hline\hline
\end{tabular}
\end{table}

Both the fat jet multiplicity and the subleading soft-drop masses, the quantities that separate the signal from $Z\to\tau^+\tau^-$, are closely reproduced by $t\bar{t}$. This is not surprising, since the decay chain $t\bar{t}\to b\bar{b}\tau^+\tau^-$ also yields 
four hard objects per event. Trained against $t\bar{t}$ alone, the event-level classifier 
reached an AUC of $0.855 \pm 0.008$, with $60.4 \pm 2.5$\% signal efficiency at 10\% 
background efficiency. The leading inputs changed accordingly, from the subleading masses to $N_\tau$ (0.289) and $E_T^{\rm miss}$ (0.204).

\subsection{Combined background}

\begin{figure}[h!]
\centering
\includegraphics[width=0.7\textwidth]
    {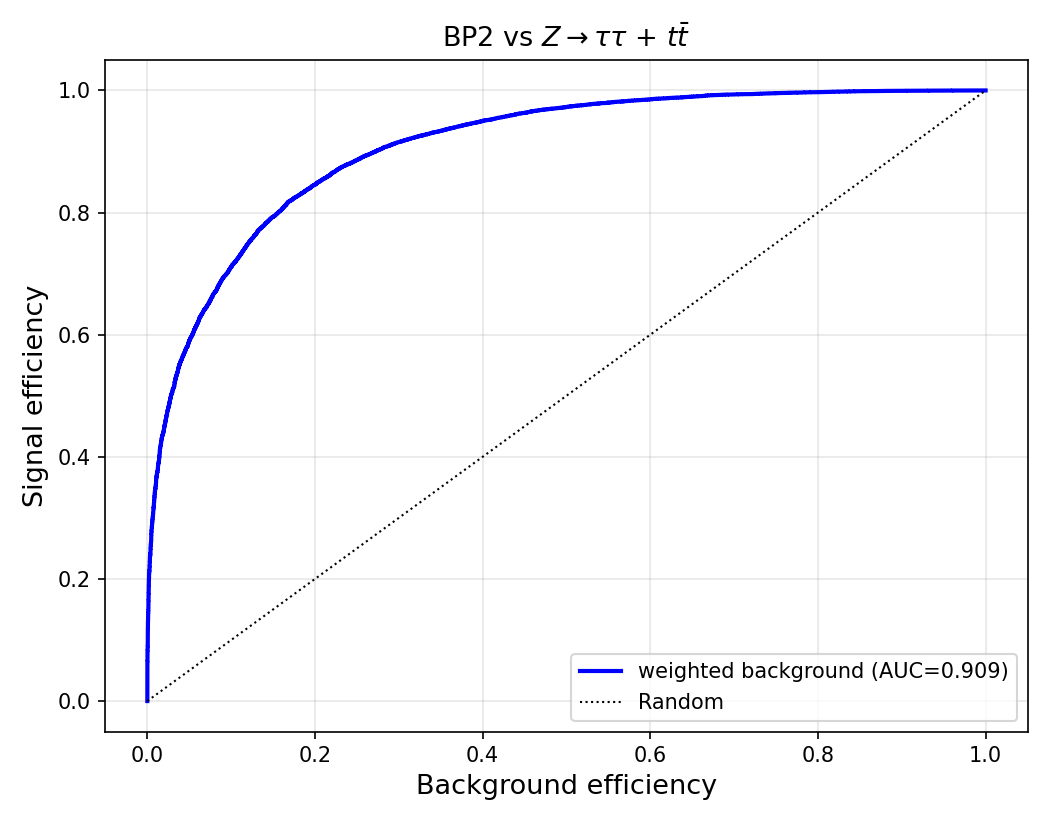}
\caption{ROC curve for the event-level classifier 
against the cross-section-weighted combination of 
$Z\to\tau^+\tau^-$ and $t\bar{t}$.}
\label{fig:rocw}
\end{figure}

For a meaningful combination, each simulated event was assigned a weight $\sigma/N$, so that the two processes enter in proportion to their cross sections: 76.4\% $Z\to\tau^+\tau^-$ and 23.6\% $t\bar{t}$. These fractions refer only to the two simulated samples before any classifier requirement, and are not the composition of the complete Standard Model background. Signal and background weights were then normalised to equal totals for training. The weighted classifier achieved an AUC of 
$0.908 \pm 0.002$, retaining $71.1 \pm 0.9$\% of signal events at 10\% background efficiency, $58.4 \pm 0.6$\% at 5\% and $34.9 \pm 1.1$\% at 1\%. The corresponding ROC curve is shown in Fig.~\ref{fig:rocw}.

The composition of the surviving background is worth noting. Although $t\bar{t}$ contributes only 23.6\% of the total background cross section, it accounts for 57.2\% of the background passing a selection at 10\% efficiency, rising to 65.7\% at 1\%. The classifier suppresses $Z\to\tau^+\tau^-$ considerably more effectively than $t\bar{t}$, and the tighter the selection, the more $t\bar{t}$ dominates what remains.

\subsection{Validation}

All quoted uncertainties were obtained from five-fold cross-validation, stratified by class so that each fold preserves the signal-to-background ratio. Complete events are the unit of splitting at event level, so no event contributes to both training and testing. The same hyperparameters and random seed are used in every fold. For the cross-section-weighted configuration the per-event weights enter both the training and the ROC computation. No thresholds are optimised per fold: the quoted efficiencies are read from the ROC curve at fixed background efficiency. Uncertainties are the standard deviations across the five folds. Table~\ref{tab:cv} collects the results for the four principal configurations.

\begin{table}[h!]
\centering
\renewcommand{\arraystretch}{1.3}
\caption{Five-fold cross-validated classifier 
performance. Uncertainties are standard deviations 
across folds.}
\label{tab:cv}
\begin{tabular}{lccc}
\hline\hline
Configuration & AUC 
  & $\varepsilon_S$ ($\varepsilon_B=10\%$) 
  & $\varepsilon_S$ ($\varepsilon_B=1\%$) \\
\hline
Single jet vs $Z$   
  & $0.759 \pm 0.002$ & $0.363 \pm 0.003$ 
  & $0.095 \pm 0.001$ \\
Event vs $Z$        
  & $0.941 \pm 0.001$ & $0.822 \pm 0.003$ 
  & $0.479 \pm 0.018$ \\
Event vs $t\bar{t}$ 
  & $0.855 \pm 0.008$ & $0.604 \pm 0.025$ 
  & $0.275 \pm 0.027$ \\
Event vs weighted   
  & $0.908 \pm 0.002$ & $0.711 \pm 0.009$ 
  & $0.349 \pm 0.011$ \\
\hline\hline
\end{tabular}
\end{table}

The larger spread for $t\bar{t}$ reflects the smaller sample: $10^4$ events against 
$5\times10^4$ for $Z\to\tau^+\tau^-$.

\subsection{Summary}

Taken together, these results indicate that the discriminating information resides principally at the level of the event rather than in the substructure of any single jet. Against $Z\to\tau^+\tau^-$ the dominant contributions came from the multiplicity of boosted objects and the masses of the subleading ones; against $t\bar{t}$, which reproduces both, discrimination relied instead on the $\tau$ multiplicity and the 
missing transverse energy. For the cross-section-weighted combination the classifier 
reached an AUC of $0.908 \pm 0.002$, with $t\bar{t}$ constituting the majority of the surviving background among the simulated samples. These are fast-simulation classification metrics and should not be interpreted as an HL-LHC discovery sensitivity.

\section{Discussion}
\label{sec:discussion}

\subsection{Why substructure fails here}

The negative result of Section~\ref{sec:observables} deserves comment, since it runs against the expectation that a light resonance decaying to a collimated pair should be identifiable through jet substructure. 

The difficulty is that both the signal and the background are two-prong objects decaying to the same final state. Boosted taggers succeed when the signal differs structurally from what surrounds it: a two-prong $W$ or Higgs jet against one-prong QCD jets, or a three-prong top jet against both. That structural contrast is absent here.

What should separate them is the parent mass. The neutrinos in $\tau$ decays prevent this: both distributions converge on $m_J \simeq 18$~GeV despite parent masses differing by an order of magnitude.

This limitation applies to any $\tau$-rich channel and is not specific to the scNMSSM. It 
suggests that mass-based identification of light resonances decaying to $\tau$ pairs requires either a mass-regression technique using the missing transverse energy, or observables constructed from the charged decay products alone, where the neutrino contribution is absent.

\subsection{Relation to earlier boosted ditau studies}

Katz, Son and Tweedie~\cite{Katz:2010iq} introduced the ditau-jet concept for a boosted 
Higgs decaying to $\tau$ pairs and demonstrated its viability against QCD backgrounds. Conte et al.~\cite{Conte:2016zpd} applied boosted ditau tagging to light NMSSM pseudoscalars produced in superpartner cascades, and Cacciapaglia et al.~\cite{Cacciapaglia:2017ofl} proposed a dedicated search for boosted ditau resonances 
below 65~GeV in composite Higgs models.

The present study differs from all three in the background against which the tagger is tested. Those analyses compared the signal primarily against QCD jets, where the two-prong structure is genuinely distinctive. Our background is boosted $Z\to\tau^+\tau^-$, which is itself a ditau jet. The performance reported here is therefore not directly comparable with theirs, and the weaker separation we obtain should not be read as a contradiction of their results.

\subsection{The role of $t\bar{t}$}

The comparison in Table~\ref{tab:means} shows that multi-object production is not unique to the signal. In $t\bar{t}\to b\bar{b}\tau^+\tau^-$ four hard objects are produced with a fat jet multiplicity of 3.36 and subleading soft-drop masses agreeing to within about 6\%.

The discrimination that remains comes from a different source. The signal produces up to eight $\tau$ leptons, of which 1.28 are reconstructed on average, against 0.61 in $t\bar{t}$; and the neutrinos from the $W$ decays and the subsequent $\tau$ decays give a harder $E_T^{\rm miss}$ spectrum, 78.7~GeV against 56.3~GeV. Together these carry 49\% of the classifier weight when trained against $t\bar{t}$.

That $t\bar{t}$ makes up 65.7\% of the surviving background at the tightest working point, despite contributing under a quarter of the background cross section, is the practical conclusion of this study. Any analysis targeting this channel should direct its rejection effort at $t\bar{t}$ rather than at $Z\to\tau^+\tau^-$.

\subsection{Limitations}

Several restrictions bound the interpretation of these results. They concern the completeness of the simulation, the choice of signal point, and the classifier itself.

First, the background is incomplete. Multijet production with misidentified $\tau$ candidates is absent, as are $W$+jets, single-top and diboson production. Fake-$\tau$ rates in particular cannot be obtained from fast simulation and require data-driven estimation.

Second, the detector description is approximate. \textsc{Delphes} parametrises the detector response rather than simulating it, and $\tau$-tagging in a dense environment is precisely the regime in which such parametrisations are least reliable. Pile-up is not modelled, and at the HL-LHC it will affect jet masses and substructure observables significantly.

Third, no trigger requirement has been imposed. With a median fat jet $p_T$ of 61~GeV, the signal is soft by the standards of existing HL-LHC trigger menus, and whether these events would be recorded at all is a question this study cannot answer.

Beyond the simulation, the signal point itself limits the conclusions. The benchmark is in tension with present experimental limits on exotic Higgs decays (Section~\ref{sec:signal}), and no absolute signal yield is quoted for that reason. The results are shape-based: the AUC values, and the composition of the background surviving a given selection, depend on the relative distributions of signal and background and not on the signal normalisation. They are therefore unchanged if the signal rate is rescaled to a branching fraction compatible with the limits. Such a rescaling, to $\text{BR}(h_1\to a_1a_1) \lesssim 12$--$18\%$, would reduce the $h_1h_1\to4a_1$ rate by a factor of thirty to seventy, since both Higgs bosons must decay exotically, and is a further reason why these results should not be read as a sensitivity estimate.

The study is also restricted to a single mass. The value $m_{a_1} = 8.2$~GeV sets the boost and hence the degree of $\tau$-pair collimation, so the balance between jet-level and event-level information may shift with $m_{a_1}$. Above the $b\bar{b}$ threshold the final state changes entirely, and the conclusions drawn here apply only to the light, $\tau$-dominated regime.

As shown in Section~\ref{sec:results}, the classifier does not generalise reliably to background softer than the sample it was trained on, so the quoted performance is specific to the kinematic region defined by the generation requirement.

Finally, the classifier was not optimised. The hyperparameters were left at their initial values, no attempt was made to select or transform the input variables, and alternative architectures were not explored. Optimisation could therefore improve the figures quoted here, although the absence of reducible backgrounds and of detector effects beyond the fast simulation acts in the opposite direction.

\section{Conclusions}
\label{sec:conclusion}

We have examined which observables separate the boosted $a_1\to\tau^+\tau^-$ topology arising from $h_1h_1 \to 4a_1$ in the scNMSSM from Standard Model processes with similar final states, taking as signal the benchmark point identified in Ref.~\cite{mtelba2026b} with $m_{a_1} = 8.2$~GeV.

The substructure observables conventionally used to identify boosted two-prong objects were found to be ineffective in this channel. Of the eleven quantities examined, the jet mass and $\tau_{21}$ ranked lowest, with separation powers of 0.02 and 0.09 respectively. The reason is that the relevant background, boosted $Z\to\tau^+\tau^-$, is itself a two-prong object decaying to $\tau$ pairs, and the neutrinos carried away in those decays displace both reconstructed masses to a common value near 18~GeV. This conclusion is not specific to the scNMSSM and applies to any search for a light resonance decaying to $\tau$ pairs against a $\tau$-rich background.

The discriminating information was instead found to reside at the level of the event. A classifier trained on individual fat jets reached an AUC of $0.759 \pm 0.002$ against $Z\to\tau^+\tau^-$, while the corresponding event-level classifier reached $0.941 \pm 0.001$. The multiplicity of boosted objects and the soft-drop masses of the subleading ones accounted for most of this difference. Truth matching supports this interpretation: 76.8\% of signal events contain at least two fat jets matched to a parent $a_1$.

Top-quark pair production limits how far this argument can be taken. The process 
$t\bar{t}\to b\bar{b}\tau^+\tau^-$ reproduces both the fat jet multiplicity and the subleading masses of the signal, and against it the event-level AUC falls to $0.855 \pm 0.008$. There the discrimination rests instead on the $\tau$ multiplicity and the missing transverse energy. For the cross-section-weighted combination of the two simulated backgrounds the classifier reached $0.908 \pm 0.002$, retaining 71\% of signal 
events at 10\% background efficiency. Although $t\bar{t}$ contributes less than a quarter of the combined background cross section, it constitutes two thirds of what survives the tightest selection considered. Cross-slice tests show that this performance holds within the kinematic range used for training but degrades on softer background, a point any experimental implementation would need to address.

These figures are classification metrics obtained from fast simulation. They are not discovery significances, and no absolute yields are quoted. Reducible backgrounds involving jet-to-$\tau$ misidentification are absent, no trigger requirement has been imposed, and pile-up is not modelled. The results establish where the discriminating information lies in this topology rather than what sensitivity an experiment would 
achieve.

Several directions follow from this work. Mass reconstruction techniques that account for the neutrino momenta, or observables built from the charged decay products alone, may recover part of the mass separation lost here. The suppression of $t\bar{t}$ merits dedicated attention, since it dominates the surviving background among the samples considered here. A quantitative sensitivity estimate will require the remaining background processes, a defined trigger strategy, a treatment of pile-up, and data-driven fake-$\tau$ rates.


\end{document}